\documentclass[sigconf]{acmart}

\renewcommand\footnotetextcopyrightpermission[1]{} 
\usepackage{enumitem}
\usepackage{multirow}
\usepackage{graphicx}
\usepackage{float}
\usepackage{subcaption}
\usepackage[ruled, lined, linesnumbered, commentsnumbered, longend]{algorithm2e}
\AtBeginDocument{%
  \providecommand\BibTeX{{%
    \normalfont B\kern-0.5em{\scshape i\kern-0.25em b}\kern-0.8em\TeX}}}

\setcopyright{acmcopyright}
\copyrightyear{2018}
\acmYear{2018}
\acmDOI{XXXXXXX.XXXXXXX}

\acmConference[KDD '24]{Make sure to enter the correct
  conference title from your rights confirmation email}{June 03--05,
  2018}{Woodstock, NY}
\acmPrice{15.00}
\acmISBN{978-1-4503-XXXX-X/18/06}

\begin{document}

\title[CA-CDSR]{Learning Partially Aligned Item Representation for \\ Cross-Domain Sequential Recommendation}



\author{Mingjia Yin}
\email{mingjia-yin@mail.ustc.edu.cn}
\orcid{0009-0005-0853-1089}
\affiliation{%
  \institution{University of Science and Technology of China \& State Key Laboratory of Cognitive Intelligence}
  \city{Hefei}
  \country{China}
}

\author{Hao Wang}
\authornote{Corresponding author.}
\email{wanghao3@ustc.edu.cn}
\orcid{0000-0001-9921-2078}
\affiliation{%
  \institution{University of Science and Technology of China \& State Key Laboratory of Cognitive Intelligence}
  \city{Hefei}
  \country{China}
}

\author{Wei Guo}
\email{guowei67@huawei.com}
\orcid{0000-0001-8616-0221}
\author{Yong Liu}
\email{liu.yong6@huawei.com}
\orcid{0000-0001-9031-9696}
\affiliation{%
  \institution{Huawei Singapore Research Center}
  \country{Singapore}
}


\author{Zhi Li}
\email{zhilizl@sz.tsinghua.edu.cn}
\affiliation{%
  \institution{ Shenzhen International Graduate
 School, Tsinghua University}
  \city{Shenzhen}
  \country{China}
}

\author{Sirui Zhao}
\email{sirui@mail.ustc.edu.cn}
\orcid{0000-0001-8103-0321}
\affiliation{%
  \institution{University of Science and Technology of China \& State Key Laboratory of Cognitive Intelligence}
  \city{Hefei}
  \country{China}
}

\author{Zhen Wang}
\email{wangzh665@mail.sysu.edu.cn}
\orcid{0000-0001-8103-0321}
\affiliation{%
  \institution{Sun Yat-Sen University}
  \city{Guangzhou}
  \country{China}
}

\author{Defu Lian}
\email{liandefu@ustc.edu.cn}
\orcid{0000-0002-3507-9607}
\affiliation{%
  \institution{University of Science and Technology of China \& State Key Laboratory of Cognitive Intelligence}
  \city{Hefei}
  \country{China}
}

\author{Enhong Chen}
\email{cheneh@ustc.edu.cn}
\orcid{0000-0002-4835-4102}
\affiliation{%
  \institution{University of Science and Technology of China \& State Key Laboratory of Cognitive Intelligence}
  \city{Hefei}
  \country{China}
}

\begin{abstract}
Cross-domain sequential recommendation (CDSR) aims to uncover and transfer users' sequential preferences across multiple recommendation domains. While significant endeavors have been made, they primarily concentrated on developing advanced transfer modules and aligning user representations using self-supervised learning techniques. However, the problem of aligning item representations has received limited attention, and misaligned item representations can potentially lead to sub-optimal sequential modeling and user representation alignment. To this end, we propose a model-agnostic framework called \textbf{C}ross-domain item representation \textbf{A}lignment for \textbf{C}ross-\textbf{D}omain \textbf{S}equential \textbf{R}ecommendation (\textbf{CA-CDSR}), which achieves sequence-aware generation and adaptively partial alignment for item representations. Specifically, we first develop a sequence-aware feature augmentation strategy, which captures both collaborative and sequential item correlations, thus facilitating holistic item representation generation. Next, we conduct an empirical study to investigate the partial representation alignment problem from a spectrum perspective. It motivates us to devise an adaptive spectrum filter, achieving partial alignment adaptively. Furthermore, the aligned item representations can be fed into different sequential encoders to obtain user representations. The entire framework is optimized in a multi-task learning paradigm with an annealing strategy. Extensive experiments have demonstrated that CA-CDSR can surpass state-of-the-art baselines by a significant margin and can effectively align items in representation spaces to enhance performance. The anonymous code can be found at \textcolor{blue}{\url{https://anonymous.4open.science/r/KDD2024-58E8/}}.
\end{abstract}




\maketitle

\section{INTRODUCTION}
\begin{figure}
    \centering
    \includegraphics[width=0.49\textwidth,height=0.25\textwidth]{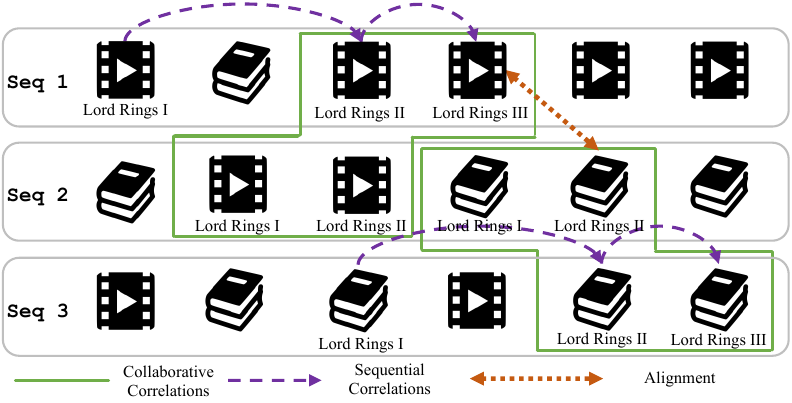}
    \caption{A toy CDSR scenario in Book and Movie domains.}
    \label{fig: motivation}
    \vspace{-0.5cm}
\end{figure}

To alleviate the ongoing situation of information explosion, recommendation systems play a dominant role, which aims to provide users with personalized services by suggesting suitable items instead of leaving them to self-seeking~\cite{wu2023survey, end4rec, hypersorec, ubcs, MCNE}. The sequential recommendation (SR) system is an important research area within the field of recommendation systems, aiming to capture users' dynamic preferences from chronologically organized interaction sequences~\cite{sequential_recommendation_survey1, sequential_recommendation_survey2, sequential_recommendation_survey3, GRU4Rec, SASRec, apgl4sr}. Despite achieving notable success, conventional methods often face challenges in effectively modeling user behavior due to the data-sparsity issue based on a single recommendation domain \cite{zhu2021survey1, zang2022survey2, he2023survey}. 



To address the issue of mining and transferring users' sequential preferences across different domains, the Cross-Domain Sequential Recommendation (CDSR) problem has been proposed and received lots of attention. Early works focused on developing sophisticated transfer modules, like $\pi$-net~\cite{pi-net} and PSJNet~\cite{psjnet} utilized gated transfer modules to transfer knowledge between domains. Subsequent studies, including MIFN~\cite{MIFN} and DA-GCN~\cite{dagcn}, leveraged the powerful high-order modeling ability of graph neural networks (GNNs) for cross-domain knowledge transfer. In recent times, the rise of self-supervised learning has led to the adoption of a contrastive learning paradigm in some works, such as C$^2$DSR~\cite{c2dsr} and Tri-CDR~\cite{Tri-CDR}, to align cross-domain user representations and facilitate effective knowledge transfer between domains.

Most existing works rely on randomly initialized item representations without elaborate processing, leading to potential misguidance in subsequent sequential modeling and user representation alignment. Therefore, we aim to align item representations for CDSR in a model-agnostic manner. Our idea is to explore the implicit item correlations across different domains. To illustrate this concept, we provide a toy scenario in Fig.~\ref{fig: motivation}, highlighting two essential processes for aligning item representations: \textbf{(1)~item representation generation}.
To explore cross-domain item correlations, it is essential to first model single-domain item correlations, i.e., collaborative and sequential correlations. For instance, in Fig. \ref{fig: motivation}, the movie "\textit{Lord Rings II}" appears in both Seq.1 and Seq.2. Analyzing the neighboring items reveals a significant connection between the movie "\textit{Lord Rings I}" in Seq.2 and the movie "\textit{Lord Rings III}" in Seq.1, representing a collaborative correlation. Additionally, we observe a sequential correlation, indicating a preference evolution trend among these three movies from the "\textit{Lord Rings}" series. \textbf{(2)~item representation alignment}. Cross-domain item representations with implicit correlations should be partially aligned. We notice that the "\textit{Lord Rings}" books exhibit similar single-domain correlations as the movies, making it reasonable to assume they could be aligned accordingly. However, the content of books and movies does not strictly correspond (inherent domain gap), meaning they should not be strictly aligned, which we define as partial alignment.
Ideally, aligned item representations can help recommend items that users have not interacted with in other domains. For instance, we could recommend the movie \textit{"Lord Rings III"} to a user who recently read the book \textit{"Lord Rings II"}, even if they have not watched movies.

There are several challenges to solving the problem of item representation alignment in CDSR to achieve this goal. First, previous works, such as~\cite{SimGCL, SGL}, have employed a self-supervised learning paradigm to model item representations and enhance their effectiveness with feature augmentation techniques~\cite{representation_augmentation_survey, SimGCL}. However, most existing 
feature augmentation methods are sequence-agnostic~\cite{representation_augmentation_survey, representation_aug_cv, SimGCL}, neglecting sequential correlations and resulting in sub-optimal item representations that can mislead the subsequent item alignment process. How to incorporate both collaborative and sequential information into item representation learning is a nontrivial problem. Secondly, achieving partial item representation alignment is a nontrivial task. The scarcity of annotated data presents challenges in determining the specific portion of the item representation that requires alignment. Previous studies\cite{c2dsr, Tri-CDR} commonly resorted to adjusting the weights of alignment losses to control the degree of user alignment. However, this approach is overly simplistic and can give rise to negative transfer issues when the alignment weight is high. Conversely, a lower weight may lead to the omission of valuable information during the transfer process. An ideal method should have the capability to adaptively identify the relevant parts of item representations for alignment. Hence, the pursuit of adaptive partial alignment across domains remains an unresolved challenge in the field of CDSR.

To address the challenges mentioned above, we propose a model-agnostic framework \textbf{CA-CDSR} named \textbf{C}ross-domain item representation \textbf{A}lignment for \textbf{C}ross-\textbf{D}omain \textbf{S}equential \textbf{R}ecommendation problem, which aims to achieve sequence-aware representation generation and partial alignment. To this end, we first introduce a sequence-aware feature augmentation strategy that condenses collaborative and sequential information into item representations, resulting in stronger representational ability for the subsequent alignment process. We then conduct an empirical study to reveal the partial representation alignment problem from a spectrum perspective. Based on this analysis, we propose a simple yet effective spectrum filter to adaptively align item representations across different domains. Additionally, we employ a graph-based sequential encoder to obtain user representations based on aligned item representations. Notably, as a model-agnostic framework, CA-CDSR can be integrated with most sequential recommenders. We optimize the framework in a multi-task learning paradigm with an annealing strategy to facilitate mutual reinforcement of each module. In summary, our contributions can be briefly summarized as follows:
\begin{itemize}[leftmargin=*]
    \item We study a novel problem of aligning item representations for CDSR, to capture implicit item correlations and achieve adaptive partial alignment in a model-agnostic manner.

    \item We propose a sequence-aware feature augmentation strategy for item representation generation. It can effectively capture collaborative and sequential item correlations.
    
    \item We conduct an innovative empirical study to verify the necessity of partial alignment of item representations across domains from a spectrum perspective and propose an Adaptive Spectrum Filter to alleviate the domain gap issues efficiently. 
    
    \item Extensive experiments have demonstrated that CA-CDSR can significantly surpass competitive baselines. Furthermore, detailed analysis has revealed the generality of CA-CDSR for alignment.
\end{itemize}

\section{RELATED WORKS}
\subsection{Sequential Recommendation System}\label{sec: related_sr}
As one of the most influential branches of recommendation systems, sequential recommendation (SR) aims to discover users' dynamic preferences in users' interaction sequences \cite{sequential_recommendation_survey1, sequential_recommendation_survey2, sequential_recommendation_survey3, dr4sr, guesr, huang2024survey}. With the rise of deep learning, researchers have made efforts to incorporate deep neural networks for modeling complex sequential preferences. These include CNN \cite{Caser, CosRec}, RNN \cite{GRU4Rec, narm}, and GNN \cite{SRGNN, GCSAN, GCEGNN}. Among all the deep SR models, Transformer-based methods, such as SASRec \cite{SASRec} and BERT4Rec \cite{bert4rec}, have gained dominance due to their powerful modeling capability. To address the data sparsity issue, Self-Supervised Learning (SSL) \cite{ssl_rec_survey} has been introduced in SR. S$^3$-Rec \cite{S3Rec} proposed auxiliary pre-training tasks based on different data views. Subsequent works, like CL4SRec \cite{CL4SRec} and ICLRec \cite{ICLRec}, employed sequence-level contrastive learning to better capture users' sequential preferences. Recent works like DuoRec \cite{DuoRec} constructed contrastive pairs at the model level.

\subsection{Cross-Domain Sequential Recommendation}\label{sec: related_cdr}
Despite the prosperity of recommendation systems, conventional methods often face the challenges of data sparsity and cold-start, as they are limited to one single recommendation domain. To address these problems, Cross-Domain Sequential Recommendation (CDSR) methods have been proposed to transfer knowledge from relatively richer domains to sparser domains \cite{zhu2021survey1, zang2022survey2, areil}.


CDSR focuses on mining and transferring users' sequential preferences across domains. Early attempts such as $\pi$-net \cite{pi-net} and PSJNet \cite{psjnet} proposed transferring domain knowledge using elaborate transfer modules. Additionally, GNNs are known for their ability to capture high-order relationships \cite{GNN_survey, RS_GNN_survey}, which motivated the development of DA-GCN \cite{dagcn} and MIFN \cite{MIFN} that utilize graphs to model the complex relationships between domains. Recently, some works proposed to learn transferable item representations with item description text, like UniSRec\cite{UniSRec} and VQ-Rec\cite{VQ-Rec}. Moreover, self-supervised learning has been adopted to align user representations in CDSR. For instance, C$^2$DSR \cite{c2dsr} proposed randomly substituting items belonging to one domain in a sequence to generate contrastive samples. Tri-CDR \cite{Tri-CDR} instead employed sequence-level contrastive learning to explore coarse-grained similarities and fine-grained distinctions among domain sequences.

However, most existing CDSR works overlook the item representation alignment problem. Misaligned item representations will potentially mislead subsequent sequential modeling and user representation alignment processes.

\section{PROBLEM DEFINITION}\label{sec: problem_definition}
In cross-domain sequential recommendation, the main task is to model user preferences based on their interaction records in different domains and recommend the next item to them in each domain. In this work, we consider a CDSR scenario consisting of two domains, where every user has interaction sequences in both domains. We can formally define the problem as follows:

\begin{definition} 
\textit{(\textbf{Cross-Domain Sequential Recommendation}). Denote the two domains as domain $\mathbf{X}$ and domain $\mathbf{Y}$. We define the user set $\mathcal{U}$, the item set $\mathcal{X}$ in domain $\mathbf{X}$, and the item set $\mathcal{Y}$ in domain $\mathbf{Y}$. For each user $u \in \mathcal{U}$, their interaction sequence in domain $\mathbf{X}$ is denoted as $S_u^{\mathbf{X}} = [x_1, x_2, \dots, x_t, \dots, x_{|S_u^{\mathbf{X}}|}]$, where $x_t \in \mathcal{X}$ is the item interacted by user $u$ at time step $t$, and $|S_u^{\mathbf{X}}|$ is the length of the sequence with a maximum value $N$. Similarly, $S_u^{\mathbf{Y}}$ can be defined for domain $\mathbf{Y}$. Furthermore, we can combine $S_u^{\mathbf{X}}$ and $S_u^{\mathbf{Y}}$ to obtain a mixed sequence $S_u = [x_1, y_1, y_2, x_2, \dots, x_{|S_u^{\mathbf{X}}|}, \dots, y_{|S_u^{\mathbf{Y}}|}]$, which is organized by timestamps. Eventually, given the sequence triplet ($S_u^{\mathbf{X}}$, $S_u^{\mathbf{Y}}$, $S_u$), the goal of CDSR is to predict the next item for each user $u \in \mathcal{U}$ in domain $\mathbf{X}$ and $\mathbf{Y}$, respectively.}
\end{definition}

Building upon the behavior sequences, our objective is to model the collaborative and sequential correlations between items. To accomplish this, we employ graph-based self-supervised learning, which has been recognized for its capacity to capture high-order correlations. Specifically, we construct two single-domain graph $\mathcal{A}^{\mathbf{X}}$ and $\mathcal{A}^{\mathbf{Y}}$ and a cross-domain graph $\mathcal{A}$ as follows:

\begin{definition} 
\textit{(\textbf{Directed Item-Item Graph}). For domain $\mathbf{X}$, we construct an item-item graph $\mathcal{A}^{\mathbf{X}} \in \mathbb{R}^{|\mathcal{X}| \times |\mathcal{X}|}$, where the matrix entry $\mathcal{A}_{ij}^{\mathbf{X}} = 1$ if $x_j$ is next to $x_i$ in any sequences in $S^{\mathbf{X}}$. Similarly, $\mathcal{A}^{\mathbf{Y}} \in \mathbb{R}^{|\mathcal{Y}| \times |\mathcal{Y}|}$ and $\mathcal{A} \in \mathbb{R}^{(|\mathcal{X}| + |\mathcal{Y}|) \times (|\mathcal{X}| + |\mathcal{Y}|)}$ can be constructed.}
\end{definition}

Then we will detail how to generate item representations enriched with collaborative and sequential information and achieve adaptively partial alignment.

\section{METHODOLOGY}
\begin{figure*}
    \centering
    \includegraphics[width=0.96\textwidth,height=0.44\textwidth]{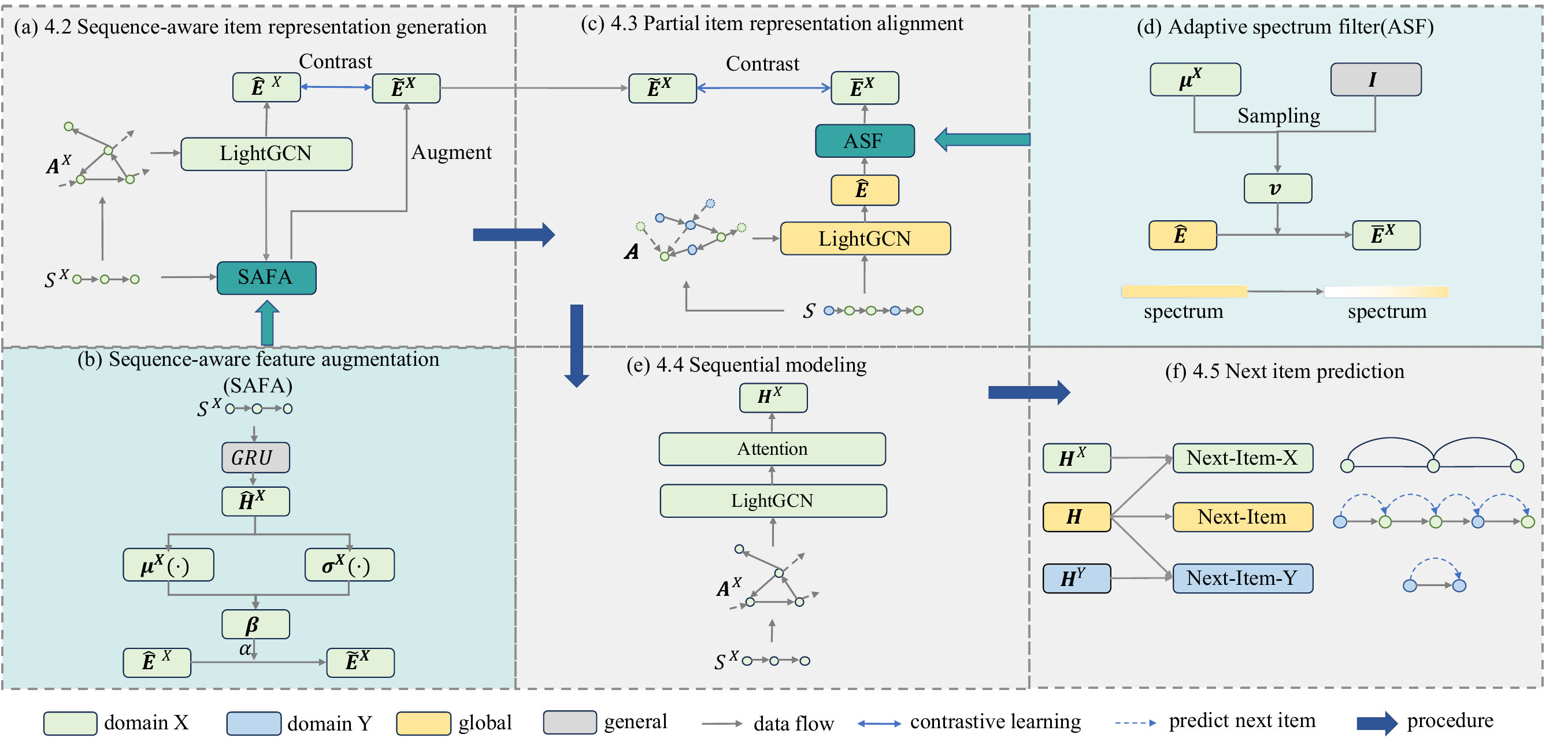}
    \caption{The framework of the proposed CA-CDSR model. We use domain $\mathbf{X}$ as an example for domain-specific modules.}
    \label{fig: framework}
\end{figure*}

\subsection{Overall Framework}

Despite the success of existing Cross-Domain Sequential Recommendation (CDSR) works, few studies have focused on the partial item representation alignment problem in CDSR, which often leads to inferior performance. To address this issue, we propose the \textbf{C}ross-domain Representation \textbf{A}lignment for \textbf{C}ross-\textbf{D}omain \textbf{S}equential \textbf{R}ecommendation (\textbf{CA-CDSR}) framework, as illustrated in Fig. \ref{fig: framework}. The CA-CDSR framework aims to achieve sequence-aware item representation modeling and partial item representation alignment for CDSR. First, in Sec. \ref{sec: intra_alignment}, we perform sequence-aware representation generation in a self-supervised learning (SSL) fashion. We introduce a novel sequence-aware feature augmentation strategy (SAFA) to preserve collaborative and sequential correlations in item representations, which serves as the foundation for the subsequent representation alignment process. Then, in Sec. \ref{sec: inter_alignment}, we investigate the partial item representation alignment problem from a spectrum perspective through an empirical study. Motivated by this study, we propose an adaptive spectrum filter (ASF) that achieves partial representation alignment adaptively and efficiently. Subsequently, in Sec. \ref{sec: sequential_modeling}, we adopt a graph-enhanced self-attention network to obtain user representations based on the aligned item representations. Finally, in Sec. \ref{sec: model_optimization}, we elaborate on the optimization of the CA-CDSR through a multi-task learning paradigm. Notably, CA-CDSR can be easily integrated with different backbone models.

\subsection{Sequence-Aware Item Representation}\label{sec: intra_alignment}

Previous methods have relied on data augmentation techniques in the discrete space, such as masking an item from a sequence. However, this approach poses challenges in generating semantically consistent augmentation samples\cite{DuoRec}.

To begin with, we take the item representation generation of the domain $\mathbf{X}$ as an example. The modeling process for representations in domain $\mathbf{Y}$ follows a similar approach. However, due to the data sparsity issue, obtaining ideal item representations without sufficient supervised signals is challenging. Therefore, we aim to model item representations in a contrastive learning fashion. Previous methods have relied on data augmentation techniques in the discrete space, such as masking an item from a sequence. However, this approach poses challenges in generating semantically consistent augmentation samples\cite{DuoRec}. Therefore, we propose a sequence-aware representation augmentation (SAFA) strategy, which operates in the continuous space, to construct augmented data that preserves both collaborative and sequential correlations.


\textbf{Collaborative correlations}. Motivated by the powerful ability of GNNs to capture high-order relationships\cite{GNN_survey}, we aim to capture collaborative correlations using single-domain graphs $\mathcal{A}^{\mathbf{X}} \in \mathbb{R}^{|\mathcal{X}| \times |\mathcal{X}|}$ and $\mathcal{A}^{\mathbf{Y}} \in \mathbb{R}^{|\mathcal{Y}| \times |\mathcal{Y}|}$ defined in Sec. \ref{sec: problem_definition}. Given $\mathcal{A}^{\mathbf{X}}$, we condense neighborhood information into item representations using a simplified graph neural network LightGCN\cite{LightGCN}:
\begin{equation}\label{eq: LGN_convolution}
    \mathbf{E}^{\mathbf{X}}_{(l)} = \text{Norm}(\mathcal{A}^{\mathbf{X}}) \mathbf{E}^{\mathbf{X}}_{(l-1)},
\end{equation}
\begin{equation}\label{eq: LGN_aggregation}
    \hat{\mathbf{E}}^{\mathbf{X}} = \frac{(\mathbf{E}^{\mathbf{X}}_{(0)} + \mathbf{E}^{\mathbf{X}}_{(1)} + \dots + \mathbf{E}^{\mathbf{X}}_{(L)})}{L + 1},
\end{equation}
where $\mathbf{E}^{\mathbf{X}}_{(0)}$ is a randomly initialized embedding table $\mathbf{E}^\mathbf{X} \in \mathbb{R}^{|\mathcal{X}| \times d}$, d is the embedding size, and $L$ is the number of layers in the graph convolution. With the graph-enhanced item representations $\hat{\mathbf{E}}^{\mathbf{X}}$, we can now incorporate collaborative correlation information into item representations through contrastive learning. Specifically, we first augment the item representations in the representation space:
\begin{equation}
    \tilde{\mathbf{E}}^{\mathbf{X}} = \hat{\mathbf{E}}^{\mathbf{X}} + \alpha \mathbf{\beta},
\end{equation}
where $\alpha$ is a constant scalar controlling the magnitude of augmentation, and $\mathbf{\beta} \in \mathbb{R}^{|\mathcal{X}| \times d}$ is a noise variable sampled from distribution $\mathcal{D}$, which will be detailed later. We then use InfoNCE\cite{CPC} to model item representations with collaborative correlations:
\begin{equation}\label{eq: intra-domain-alignment}
    \mathcal{L}^{\mathbf{X}} = - \sum_{i=1}^{|B|} \log{\frac
        {e^{cos(\hat{\mathbf{e}}_i^{\mathbf{X}}, \tilde{\mathbf{e}}_i^{\mathbf{X}}) / \tau}}
        {\sum_{j=1}^{|B|} e^{cos(\hat{\mathbf{e}}_i^{\mathbf{X}}, \tilde{\mathbf{e}}_j^{\mathbf{X}}) / \tau}}
    },
\end{equation}
where $|B|$ is the batch size, $cos(\cdot, \cdot)$ denotes cosine similarity, and $\tau$ is a temperature hyper-parameter. By utilizing Equation \ref{eq: intra-domain-alignment}, we can effectively capture the intra-domain collaborative item correlations.

\textbf{Sequential correlations}. Considering the collaborative correlations alone are inadequate to achieve holistic representation modeling, we further explore the sequential item correlations, which are crucial for capturing users' evolving preferences in subsequent sequence modeling. Previous feature augmentation methods\cite{SimGCL} sample noise from a data-independent distribution. However, item representations exhibit a trend of maintaining sequential correlations with other items in the sequences. It is crucial for augmented item representations to preserve these correlations. Therefore, we introduce a learnable noise generator that produces sequence-aware noise, facilitating the modeling of sequential correlations.

First, we utilize an encoder to extract sequential information. Given an interaction sequence $S^{\mathbf{X}} = [x_1, x_2, \dots, x_t, \dots, x_{|S^{\mathbf{X}}|}]$, we generate its corresponding sequential representation:
\begin{equation}
    \hat{\mathbf{h}}^{\mathbf{X}} = \text{Encoder}(S^{\mathbf{X}}),
\end{equation}
where $\text{Encoder}$ is a sequential encoder, and we instantiate it as a gated recurrent network for its effectiveness in capturing sequential signals. To capture sequential information, we define the noise distribution $\mathcal{D}$ as a learnable Multivariate Normal distribution $\mathcal{N}(\mathbf{\mu}^{\mathbf{X}}(\hat{\mathbf{h}}^{\mathbf{X}}), (\mathbf{\sigma}^{\mathbf{X}}(\hat{\mathbf{h}}^{\mathbf{X}}))^2)$. Specifically, $\mathbf{\mu}(\cdot)$ and $\mathbf{\sigma}(\cdot)$ are implemented as MLP networks, enabling the noise to be sampled from a sequence-aware distribution. Since the sampling procedure is non-differentiable, we utilize the reparameterization\cite{VAE} trick:
\begin{equation}
    \mathbf{\beta} = {\mathbf{\mu}}^{\mathbf{X}}(\hat{\mathbf{h}}^{\mathbf{X}}) + \epsilon \mathbf{\sigma}^{\mathbf{X}}(\hat{\mathbf{h}}^{\mathbf{X}}),
\end{equation}
where $\epsilon \in \mathbb{R}^{d}$ is randomly sampled from $\mathcal{N}(\mathbf{0}, \mathbf{I})$.

By modeling both collaborative and sequential item correlations, we obtain item representations enriched with holistic knowledge, which lay a solid foundation for the subsequent alignment process.

\subsection{Partial Item Representation Alignment}\label{sec: inter_alignment}
Given the item representations with well-preserved collaborative and sequential correlations, we can now start to perform representation alignment across domains. However, we are still in the dilemma of lacking explicit data to guide the alignment process. Therefore, we instead perform item representation alignment based on common users in both two domains.

To leverage the information of these common users, we first concatenate the two embedding matrices $\mathbf{E}^{\mathbf{X}} \in \mathbb{R}^{|\mathcal{X}| \times d}$ and $\mathbf{E}^{\mathbf{Y}} \in \mathbb{R}^{|\mathcal{Y}| \times d}$ into $\mathbf{E} \in \mathbb{R}^{(|\mathcal{X}| + |\mathcal{Y}|) \times d}$. Then, given the mixed item-item graph $\mathcal{A} \in \mathbb{R}^{(|\mathcal{X}| + |\mathcal{Y}|) \times (|\mathcal{X}| + |\mathcal{Y}|)}$ in Sec. \ref{sec: problem_definition}, which is constructed based on common users, we can obtain the cross-domain global item representation $\hat{\mathbf{E}}$ as in Eq.(~\ref{eq: LGN_convolution}) and Eq.(~\ref{eq: LGN_aggregation}). With the cross-domain global item representation $\hat{\mathbf{E}}$, and the augmented local item representations $\tilde{\mathbf{E}}^{\mathbf{X}}$ and $\tilde{\mathbf{E}}^{\mathbf{Y}}$, we can now perform representation alignment by pulling in local and global item representations:
\begin{equation}\label{eq: inter-domain-alignment}
    \mathcal{L}_{\text{align}}^{\text{inter}} = - \sum_{i=1}^{|B|} \log{\frac
        {e^{\cos(\hat{\mathbf{e}}_i, \tilde{\mathbf{e}}_i^{\mathbf{X}}) / \tau}}
        {\sum_{j=1}^{|B|} e^{\cos(\hat{\mathbf{e}}_i, \tilde{\mathbf{e}}_j^{\mathbf{X}}) / \tau}}
    } - \sum_{i=1}^{|B|} \log{\frac
        {e^{\cos(\hat{\mathbf{e}}_i, \tilde{\mathbf{e}}_i^{\mathbf{Y}}) / \tau}}
        {\sum_{j=1}^{|B|} e^{\cos(\hat{\mathbf{e}}_i, \tilde{\mathbf{e}}_j^{\mathbf{Y}}) / \tau}}
    }.
\end{equation}
Since cross-domain information is condensed in $\hat{\mathbf{E}}$, aligning $\hat{\mathbf{E}}$ with $\tilde{\mathbf{E}}^{\mathbf{X}}$ or $\tilde{\mathbf{E}}^{\mathbf{Y}}$ will facilitate the implicit alignment between item representations of domain $\mathbf{X}$ and domain $\mathbf{Y}$.

\begin{figure}[t!]
    \centering
    \begin{subfigure}[t]{0.23\textwidth}
           \centering
           \includegraphics[width=\textwidth]{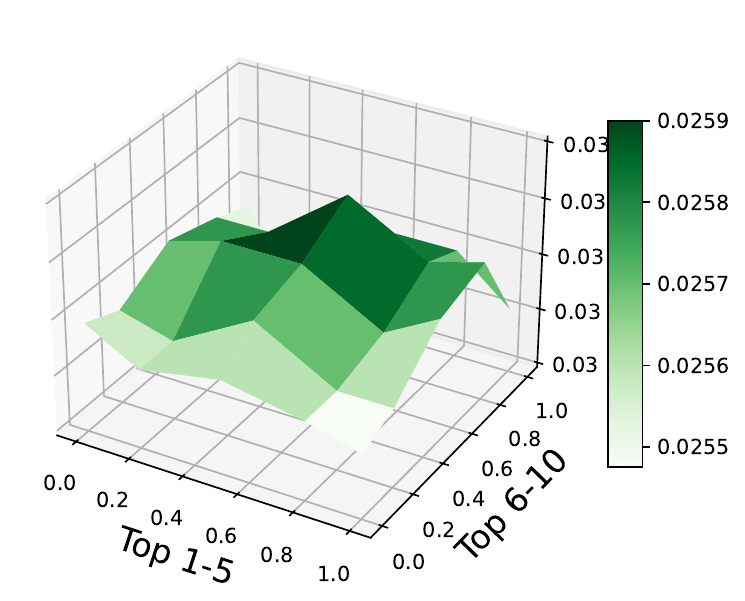}
            \caption{Food}
            \label{fig: food_empirical}
    \end{subfigure}
    \begin{subfigure}[t]{0.23\textwidth}
            \centering
            \includegraphics[width=\textwidth]{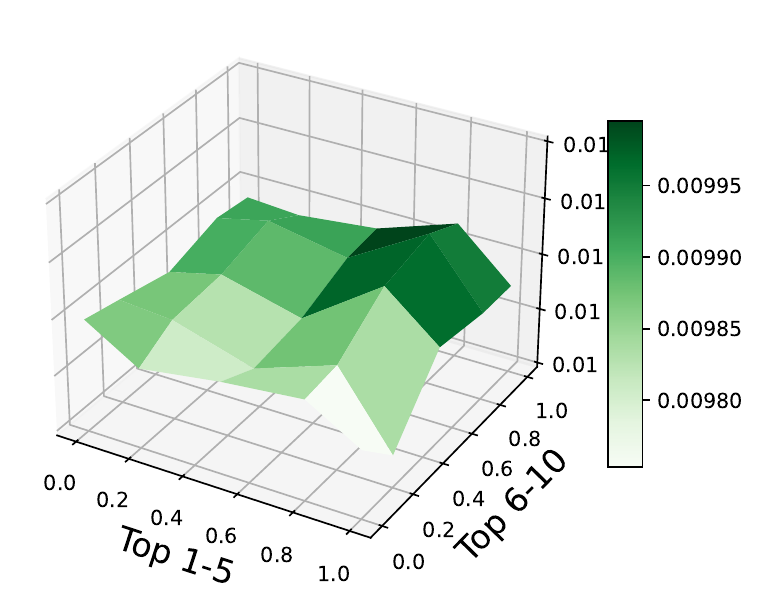}
            \caption{Kitchen}
            \label{fig: kitchen_empirical}
    \end{subfigure}
    \caption{The empirical study on manipulating the spectrum of global item representation $\hat{\mathbf{e}}_i$ on the Food-Kitchen dataset. Specifically, we focused on the top 10 largest singular values, dividing them into two groups: top 1-5 and top 6-10. We reduced the singular values in each group by multiplying a coefficient. This can be viewed as a manual filtering operation that selectively removes a portion of the information.}
    \label{fig: empirical_study}    
    \vspace{-0.2cm}
\end{figure}

 However, perfect representation alignment will eliminate the inherent domain gap between domains, leading to the negative transfer problem \cite{negative_transfer_survey, zhu2021survey1}. Therefore, we question whether the two domains should be \textit{\textbf{partially}} aligned, i.e., only a portion of the global representation $\hat{\mathbf{e}}_i$ should be aligned. Since the representation spectrum (singular values) can reflect the information carried by representations, we investigate the problem of partial item representation alignment from a spectrum perspective. Furthermore, as different singular values carry distinct information, we conduct an empirical study to examine how singular values contribute to representation alignment, as illustrated in Fig. \ref{fig: empirical_study}. From the figure, we draw the following conclusions: (1) The optimal filtering strengths for the two groups of singular values differ, indicating that different singular values exhibit varying contributions to the representation alignment process. (2) Similarly, the optimal filtering strengths differ for distinct domains, indicating that different domains require different filtering strengths. \textbf{Both of the two observations suggest the need for adaptive manipulation of the spectrum to adaptively achieve partial alignment for each domain}.

Although manual manipulation of the representation spectrum can partially mitigate the issue stemming from perfect representation alignment, two challenges remain unresolved. (1) The SVD decomposition is not easily parallelizable, resulting in inefficient training\cite{rank-1-update}. (2) The search space for manual filtering is intractable, making it impractical to identify the optimal filter for every singular value and domain. Building on empirical findings, we introduce an adaptive spectrum filter to tackle these challenges.

Regarding the first challenge, the rank-1 update method\cite{rank-1-update} enables efficient spectrum modification. However, it lacks sensitivity to domain information, and its filtering procedure is non-learnable, which fails to fulfill the aforementioned requirement. Consequently, we introduce a straightforward yet highly effective module called the adaptive spectrum filter (ASF). This enables adaptive and efficient spectrum filtering, leading to partial representation alignment. Subsequently, we provide a detailed description of the procedure, using domain $\mathbf{X}$ as an illustrative example.


Specifically, our objective is to derive the filtered global representations $\bar{\mathbf{E}}^{\mathbf{X}}$ by utilizing the cross-domain item representations $\hat{\mathbf{E}}$. Initially, we randomly sample a spectrum filter $\mathbf{v}$ from a filter distribution $\mathcal{N}(\mu^{\mathbf{X}}, 1)$. Here, $\mu^{\mathbf{X}}$ represents a learnable embedding that can be optimized using a technique akin to the reparameterization trick\cite{VAE}. Subsequently, the filtered representation can be obtained by improving the rank-1 update process in \cite{rank-1-update} as follows:
\begin{equation}
\begin{gathered}
    \hat{\mathbf{v}} = \mathbf{\hat{E}}^T \mathbf{\hat{E}} \mathbf{v}, \\
    \bar{\mathbf{E}}^{\mathbf{X}} = \bar{\mathbf{E}} - \bar{\mathbf{E}} \frac{\hat{\mathbf{v}} \hat{\mathbf{v}}^T}{\parallel \hat{\mathbf{v}} \parallel _2^2}.
\end{gathered}
\end{equation}
Similarly, we can obtain the filtered global representations $\bar{\mathbf{E}}^{\mathbf{Y}}$ for domain $\mathbf{Y}$. Now Eq. \ref{eq: inter-domain-alignment} can be re-written as:
\begin{equation}\label{eq: inter-domain-alignment-filtered}
    \mathcal{L}^{\text{inter}}_{\text{align}} = - \sum_{i=1}^{|B|} \log{\frac
        {e^{cos(\bar{\mathbf{e}}_i^{\mathbf{X}}, \tilde{\mathbf{e}}_i^{\mathbf{X}}) / \tau}}
        {\sum_{j=1}^{|B|} e^{cos(\bar{\mathbf{e}}_i^{\mathbf{X}}, \tilde{\mathbf{e}}_j^{\mathbf{X}}) / \tau}}
    } - \sum_{i=1}^{|B|} \log{\frac
        {e^{cos(\bar{\mathbf{e}}_i^{\mathbf{Y}}, \tilde{\mathbf{e}}_i^{\mathbf{Y}}) / \tau}}
        {\sum_{j=1}^{|B|} e^{cos(\bar{\mathbf{e}}_i^{\mathbf{Y}}, \tilde{\mathbf{e}}_j^{\mathbf{Y}}) / \tau}}
    }.
\end{equation}


As the filter is learnable, we can attain adaptive filtering magnitudes for various singular values. Additionally, distinct filter distributions are employed for each domain, ensuring adaptiveness to specific domains. Thus, by utilizing the adaptive spectrum filter, the representation alignment process can be appropriately guided.

\subsection{Sequential Modeling}\label{sec: sequential_modeling}

Once we have acquired the appropriately aligned item representations, we employ a sequential encoder to generate user representations. Given the impressive advancements of GNN-based sequential models and the self-attention mechanism\cite{transformer}, we employ a GNN-enhanced attentive network (GNN-Att). Importantly, as the two proposed modules are model-agnostic, alternative sequential encoders are also viable, as demonstrated in Sec. \ref{sec: versatility}.


Specifically, using the aligned item representations $\mathbf{E}^{\mathbf{X}}$, $\mathbf{E}^{\mathbf{Y}}$, and $\mathbf{E}$, we leverage LightGCN\cite{LightGCN} to generate graph-enhanced representations $\hat{\mathbf{E}}^{\mathbf{X}}$, $\hat{\mathbf{E}}^{\mathbf{Y}}$, and $\hat{\mathbf{E}}$ following Eq. \ref{eq: LGN_convolution} and Eq. \ref{eq: LGN_aggregation}. We project an interaction sequence $S_u = [x_1, y_1, y_2, x_2, \dots, x_{|S_u^{\mathbf{X}}|}, \dots, y_{|S_u^{\mathbf{Y}}|}]$ into a sequence of representations $[\hat{\mathbf{e}}^{\mathbf{X}}_1, \hat{\mathbf{e}}^{\mathbf{Y}}_1, \hat{\mathbf{e}}^{\mathbf{Y}}_2, \hat{\mathbf{e}}^{\mathbf{X}}_2, \dots, \hat{\mathbf{e}}^{\mathbf{X}}_{|S_u^{\mathbf{X}}|}, \dots, \hat{\mathbf{e}}^{\mathbf{Y}}_{|S_u^{\mathbf{Y}}|}]$, which is enhanced with graph information. We then apply positional encoding to the sequence representations, obtaining the initial input $\mathbf{H}^{(0)}$. Similar to SASRec\cite{SASRec}, the attentive network comprises multiple multi-head self-attention (MHSA) layers and feed-forward layers (FFN), which can be represented as:
\begin{equation}\label{eq: SASRec}
    \mathbf{H}^{(l)} = FFN(MHSA(\mathbf{H}^{(l - 1)})),
\end{equation}
where $\mathbf{H}^{(l)}$ is the representation in the $l$-th layer, and we denote $\mathbf{H}$ as the output user representation. Besides, we can pad $S_u$ to yield single-domain sequences $S_u^{\mathbf{X}} = [x_1, \text{[P]}, \text{[P]}, x_2, \dots, x_{|S_u^{\mathbf{X}}|}, \dots, \text{[P]}]$ and $S_u^{\mathbf{Y}} = [\text{[P]}, y_1, y_2, \text{[P]}, \dots, \text{[P]}, \dots, y_{|S_u^{\mathbf{Y}}|}]$, where $\text{[P]}$ will be projected into a zero constant vector. In this way, we can obtain single-domain user representations $\mathbf{H}^{\mathbf{X}}$ and $\mathbf{H}^{\mathbf{Y}}$ similarly as in Eq.(\ref{eq: SASRec}).

\subsection{Model Optimization}\label{sec: model_optimization}
\subsubsection{\textbf{Next-item prediction objective}}
With the generated user representations $\mathbf{H}^{\mathbf{X}}$, $\mathbf{H}^{\mathbf{Y}}$ and $\mathbf{H}$, the entire model can be optimized by the next-item prediction losses. In this paper, we adopt two recommendation objectives, whose effectiveness has been verified\cite{c2dsr}.

\textbf{Single-domain next-item prediction}. Denote the padded sequences in domain $\mathbf{X}$ as $S_u^{\mathbf{X}} = [x_1, \text{[P]}, x_2, \dots, x_{t}]$, where $\text{[P]}$ is a padding, and $x_{t + 1}$ is the next item. As illustrated in Fig. \ref{fig: framework}(f), the single-domain prediction loss of domain $\mathbf{X}$ is defined as:
\begin{equation}\label{eq: next-item-prediction-X}
\begin{gathered}
    \mathcal{L}_{\text{single-rec}}^{\mathbf{X}} = \sum_{S_u^{\mathbf{X}} \in S^{\mathbf{X}}} \sum_{t} \mathcal{L}_{\text{single}}^{\mathbf{X}} (S_u^{\mathbf{X}}, t), \\
    \mathcal{L}_{\text{single}}^{\mathbf{X}} (S_u^{\mathbf{X}}, t) = -\log \left(\text{Softmax}((\mathbf{h}_t^{\mathbf{X}} + \mathbf{h}_t) \mathbf{W}^{\mathbf{X}})_{x_{t+1}}\right),
\end{gathered}
\end{equation}
where $\mathbf{h}_t^{\mathbf{X}}$ and $ \mathbf{h}_t$ are the user representations of user $u$ at timestamp $t$, and $\mathbf{W}^{\mathbf{X}} \in \mathbb{R}^{d \times |\mathcal{X}|}$ is a learnable parameter matrix for prediction. The purpose of Eq.(\ref{eq: next-item-prediction-X}) is to maximize the probability of recommending the expected next item $x_{t+1}$ based on the user representations obtained from the single-domain sequence and mixed sequence. $\mathcal{L}_{\text{single-rec}}^{\mathbf{Y}}$ can be defined similarly.

\textbf{Cross-domain next-item prediction}. In the mixed sequences, users' sequential behaviors in one domain may also lead to an action in the other domain (Someone who has recently purchased food requiring special handling may lead to the corresponding kitchen utensils). Therefore, as depicted in Fig. \ref{fig: framework}(f), a cross-domain next-item prediction loss is defined as follows:
\begin{equation}\label{eq: next-item-prediction-cross}
\begin{gathered}
    \mathcal{L}_{\text{cross-rec}} = \sum_{S_u \in S} \sum_{t} \mathcal{L}_{\text{cross}} (S_u, t), \\
    \mathcal{L}_{\text{cross}} (S_u, t) = \begin{cases}
        -\log \left(\text{Softmax}(\mathbf{h}_t \mathbf{W}^{\mathbf{X}})_{x_{t+1}}\right) \\
        -\log \left(\text{Softmax}(\mathbf{h}_t \mathbf{W}^{\mathbf{Y}})_{y_{t+1}}\right)
    \end{cases}.
\end{gathered}
\end{equation}
Therefore, the mixed user representations $H$ can maintain recommendation power for both domains, rather than be dominated by only one single domain.

\subsubsection{\textbf{Overall learning objective}}
We now introduce the overall learning objective by combining the previous losses from Eq.(\ref{eq: intra-domain-alignment}), Eq.(\ref{eq: inter-domain-alignment-filtered}), Eq.(\ref{eq: next-item-prediction-X}), and Eq.(\ref{eq: next-item-prediction-cross}). Intuitively, during the initial training phase, the model should focus primarily on intra-domain modeling. As the representations become enriched with sufficient single-domain knowledge, the framework can progressively shift its focus toward capturing more intricate cross-domain relationships. Based on this rationale, we present an annealing strategy to achieve a balance between single-domain and cross-domain training as follows:
\begin{equation}\label{eq: total loss}
\begin{gathered}
    \mathcal{L}_{rec} = \eta (\mathcal{L}_{\text{intra-domain}}) + (1 - \eta) \mathcal{L}_{\text{inter-domain}}, \\
    \mathcal{L}_{\text{intra-domain}} = \mathcal{L}_{\text{single-rec}}^{\mathbf{X}} + \mathcal{L}_{\text{single-rec}}^{\mathbf{Y}} + \lambda_1 (\mathcal{L}^{\mathbf{X}} + \mathcal{L}^{\mathbf{Y}}), \\
    \mathcal{L}_{\text{inter-domain}} = \mathcal{L}_{\text{cross-rec}} + \lambda_2 \mathcal{L}_{\text{align}}^{\text{inter}},
\end{gathered}
\end{equation}
where $\eta$ is the annealing factor, which is initialized as 1 and linearly decreased to 0.5 with $N_{anneal}$ annealing steps, $\lambda_1$ is a hyper-parameter controlling sequence-aware representation modeling, and $\lambda_2$ is a hyper-parameter controlling partial representation alignment. The entire framework will be optimized in a multi-task learning paradigm, facilitating mutual reinforcement of each module.

\subsubsection{\textbf{Computational complexity}}
The computational complexity of CA-CDSR primarily arises from the graph convolution operation in Eq.(\ref{eq: LGN_convolution}). To maintain brevity, we will focus solely on discussing its complexity. By utilizing sparse matrix multiplication operations, the complexity of Eq.(\ref{eq: LGN_convolution}) is $O(|\mathcal{E}|d)$, where $|\mathcal{E}|$ represents the total number of interaction records. By selecting $d << |\mathcal{E}|$, CA-CDSR achieves linear complexity relative to the number of interaction records within the datasets.

\section{EXPERIMENTAL EVALUATION}
\subsection{Experimental Settings}
\subsubsection{\textbf{Datasets}}

\begin{table}[]
\caption{Statistics of the datasets.}
\label{tab: datasets}
\resizebox{\linewidth}{!}{
\begin{tabular}{c|ccccc}
\hline
Scenario      & \#Item & \#Train                  & \#Val & \#Test & Avg. Length            \\ \hline
Food          & 29,207 & \multirow{2}{*}{34,117}  & 2,722 & 2,747  & \multirow{2}{*}{9.91}      \\
Kitchen       & 34,886 &                          & 5,451 & 5,659  &                         \\ \hline
Movie         & 36,845 & \multirow{2}{*}{58,515}  & 2,032 & 1,978  & \multirow{2}{*}{11.98}          \\
Book          & 63,937 &                          & 5,612 & 5,730  &  \\ \hline
Entertain & 8,367  & \multirow{2}{*}{120,635} & 4,525 & 4,485  & \multirow{2}{*}{29.94}    \\
Education     & 11,404 &                          & 2,404 & 2,300  &                       \\ \hline
\end{tabular}
}
\end{table}
To verify the superiority of the proposed CA-CDSR, we follow C$^2$DSR\cite{c2dsr} to construct CDSR scenarios based on two widely-adopted public datasets \textbf{Amazon}\footnote{\url{http://jmcauley.ucsd.edu/data/amazon/index_2014.html}} and \textbf{HVIDEO}\footnote{\url{https://bitbucket.org/Catherine_Ma/pinet_sigir2019/src/master/HVIDEO/}}. Specifically, we build three CDSR scenarios "Food-Kitchen" (Amazon), "Movie-Book" (Amazon), and "Entertain-Education" (HVIDEO). As for preprocessing the datasets, we first extract common users belonging to both domains and filter out users and items whose number of interactions is fewer than 10. Then we ensure the interaction sequences for each domain contain at least 3 items. The statistics of the preprocessed datasets are summarized in Table \ref{tab: datasets}.

\subsubsection{\textbf{Compared Baselines}}
\begin{table*}[]
    \centering
    \caption{The overall performance of all methods. The best results are bolded while the second-best results are underlined in each row. * means improvements are statistically significant with p<0.05, and ** means p<0.01.}
    \label{tab: overall performance}
    \renewcommand\arraystretch{0.9}
    \vspace{-0.2cm}
    \resizebox{\textwidth}{!}{
    \begin{tabular}{l|l|cc|cc|ccc|cccc|cc}
\hline
Datasets                       & Metrics   & BPR    & ItemKNN & NCF    & CoNet  & GRU4Rec & SASRec & SR-GNN & $\pi$-net & MIFN   & C2DSR  & Tri-CDR         & CA-CDSR                & Improv \\ \hline
\multirow{5}{*}{Food}          & MRR       & 0.0101 & 0.0097  & 0.0111 & 0.0102 & 0.0143  & 0.0176 & 0.0193 & 0.0189    & 0.0210 & 0.0247 & \underline{0.0254} & \textbf{0.0271$^{**}$} & 6.69\% \\
                               & Recall@10 & 0.0174 & 0.0171  & 0.0195 & 0.0182 & 0.0265  & 0.0337 & 0.0361 & 0.0352    & 0.0385 & 0.0461 & \underline{0.0468} & \textbf{0.0506$^{**}$} & 8.12\% \\
                               & Recall@20 & 0.0252 & 0.0251  & 0.0285 & 0.0256 & 0.0382  & 0.0475 & 0.0523 & 0.0497    & 0.0551 & 0.0626 & \underline{0.0639} & \textbf{0.0716$^{**}$} & 12.1\% \\
                               & NDCG@10   & 0.0124 & 0.0115  & 0.0134 & 0.0125 & 0.0172  & 0.0221 & 0.0235 & 0.0220    & 0.0244 & 0.0273 & \underline{0.0284} & \textbf{0.0303$^{**}$} & 6.69\% \\
                               & NDCG@20   & 0.0143 & 0.0135  & 0.0159 & 0.0148 & 0.0205  & 0.0263 & 0.0279 & 0.0257    & 0.0286 & 0.0316 & \underline{0.0327} & \textbf{0.0356$^{**}$} & 8.87\% \\ \hline
\multirow{5}{*}{Kitchen}       & MRR       & 0.0040 & 0.0037  & 0.0045 & 0.0041 & 0.0064  & 0.0072 & 0.0075 & 0.0069    & 0.0082 & 0.0093 & \underline{0.0098} & \textbf{0.0105$^{**}$} & 7.14\% \\
                               & Recall@10 & 0.0052 & 0.0050  & 0.0058 & 0.0060 & 0.0088  & 0.0112 & 0.0117 & 0.0106    & 0.0126 & 0.0145 & \underline{0.0150} & \textbf{0.0164$^{**}$} & 9.33\% \\
                               & Recall@20 & 0.0075 & 0.0071  & 0.0082 & 0.0085 & 0.0125  & 0.0158 & 0.0166 & 0.0150    & 0.0173 & 0.0209 & \underline{0.0212} & \textbf{0.0233$^{*}$}  & 9.91\% \\
                               & NDCG@10   & 0.0039 & 0.0037  & 0.0041 & 0.0042 & 0.0060  & 0.0074 & 0.0078 & 0.0076    & 0.0081 & 0.0097 & \underline{0.0100} & \textbf{0.0107$^{**}$} & 7.00\% \\
                               & NDCG@20   & 0.0046 & 0.0043  & 0.0049 & 0.0049 & 0.0070  & 0.0088 & 0.0091 & 0.0089    & 0.0094 & 0.0114 & \underline{0.0117} & \textbf{0.0125$^{**}$} & 6.84\% \\ \hline \hline
\multirow{5}{*}{Movie}         & MRR       & 0.0069 & 0.0068  & 0.0074 & 0.0073 & 0.0088  & 0.0091 & 0.0093 & 0.0100    & 0.0118 & 0.0171 & \underline{0.0179} & \textbf{0.0197$^{**}$} & 10.1\% \\
                               & Recall@10 & 0.0091 & 0.0095  & 0.0098 & 0.0099 & 0.0126  & 0.0124 & 0.0123 & 0.0138    & 0.0171 & 0.0258 & \underline{0.0265} & \textbf{0.0296$^{*}$}  & 11.7\% \\
                               & Recall@20 & 0.0115 & 0.0120  & 0.0123 & 0.0125 & 0.0155  & 0.0152 & 0.0157 & 0.0170    & 0.0221 & 0.0309 & \underline{0.0314} & \textbf{0.0372$^{**}$} & 18.5\% \\
                               & NDCG@10   & 0.0070 & 0.0071  & 0.0075 & 0.0077 & 0.0078  & 0.0075 & 0.0081 & 0.0089    & 0.0117 & 0.0185 & \underline{0.0192} & \textbf{0.0207$^{*}$}  & 7.81\% \\
                               & NDCG@20   & 0.0077 & 0.0079  & 0.0083 & 0.0082 & 0.0084  & 0.0082 & 0.0090 & 0.0093    & 0.0126 & 0.0199 & \underline{0.0203} & \textbf{0.0226$^{**}$} & 11.3\% \\ \hline
\multirow{5}{*}{Book}          & MRR       & 0.0037 & 0.0036  & 0.0041 & 0.0041 & 0.0048  & 0.0053 & 0.0050 & 0.0063    & 0.0073 & 0.0095 & \underline{0.0098} & \textbf{0.0104$^{**}$} & 6.12\% \\
                               & Recall@10 & 0.0055 & 0.0052  & 0.0061 & 0.0057 & 0.0072  & 0.0081 & 0.0078 & 0.0096    & 0.0111 & 0.0162 & \underline{0.0163} & \textbf{0.0173$^{**}$} & 6.13\% \\
                               & Recall@20 & 0.0067 & 0.0062  & 0.0070 & 0.0068 & 0.0084  & 0.0095 & 0.0095 & 0.0116    & 0.0128 & 0.0179 & \underline{0.0186} & \textbf{0.0206$^{**}$} & 10.8\% \\
                               & NDCG@10   & 0.0043 & 0.0040  & 0.0044 & 0.0045 & 0.0047  & 0.0055 & 0.0056 & 0.0057    & 0.0062 & 0.0105 & \underline{0.0109} & \textbf{0.0116$^{*}$}  & 6.42\% \\
                               & NDCG@20   & 0.0055 & 0.0047  & 0.0053 & 0.0054 & 0.0058  & 0.0067 & 0.0064 & 0.0069    & 0.0074 & 0.0112 & \underline{0.0116} & \textbf{0.0123$^{*}$}  & 6.03\% \\ \hline \hline
\multirow{5}{*}{Entertainment} & MRR       & 0.3244 & 0.3340  & 0.3204 & 0.3229 & 0.3252  & 0.3523 & 0.3539 & 0.3718    & -      & 0.4521 & \underline{0.4594} & \textbf{0.4724$^{**}$} & 2.83\% \\
                               & Recall@10 & 0.4550 & 0.4581  & 0.4458 & 0.4641 & 0.4673  & 0.5094 & 0.5025 & 0.5128    & -      & 0.6193 & \underline{0.6217} & \textbf{0.6320$^{**}$} & 1.66\% \\
                               & Recall@20 & 0.5009 & 0.5199  & 0.5059 & 0.5109 & 0.5041  & 0.5495 & 0.5588 & 0.5645    & -      & 0.6810 & \underline{0.6840} & \textbf{0.6957$^{**}$} & 1.71\% \\
                               & NDCG@10   & 0.3940 & 0.4008  & 0.4238 & 0.4514 & 0.4419  & 0.4586 & 0.4601 & 0.4432    & -      & 0.4889 & \underline{0.4927} & \textbf{0.5062$^{**}$} & 2.74\% \\
                               & NDCG@20   & 0.4170 & 0.4330  & 0.4485 & 0.4540 & 0.4535  & 0.4903 & 0.4919 & 0.4740    & -      & 0.4984 & \underline{0.5012} & \textbf{0.5195$^{**}$} & 3.65\% \\ \hline
\multirow{5}{*}{Education}     & MRR       & 0.2890 & 0.3024  & 0.2904 & 0.3129 & 0.3225  & 0.3407 & 0.3438 & 0.3457    & -      & 0.4012 & \underline{0.4058} & \textbf{0.4174$^{**}$} & 2.86\% \\
                               & Recall@10 & 0.4374 & 0.4239  & 0.4453 & 0.4497 & 0.4940  & 0.5206 & 0.5246 & 0.5003    & -      & 0.5834 & \underline{0.5861} & \textbf{0.5957$^{**}$} & 1.64\% \\
                               & Recall@20 & 0.4651 & 0.4694  & 0.4684 & 0.4980 & 0.5470  & 0.5593 & 0.5751 & 0.5542    & -      & 0.6422 & \underline{0.6451} & \textbf{0.6562$^{**}$} & 1.72\% \\
                               & NDCG@10   & 0.3493 & 0.3539  & 0.3697 & 0.3747 & 0.4244  & 0.4216 & 0.4236 & 0.4155    & -      & 0.4411 & \underline{0.4449} & \textbf{0.4547$^{*}$}  & 2.20\% \\
                               & NDCG@20   & 0.3587 & 0.3633  & 0.3835 & 0.3847 & 0.4270  & 0.4420 & 0.4380 & 0.4354    & -      & 0.4558 & \underline{0.4593} & \textbf{0.4696$^{**}$} & 2.24\% \\ \hline
\end{tabular}
    }
\end{table*}

We compare the performance of our method with four groups of state-of-the-art baselines:
\begin{itemize}[leftmargin=*]
    \item \textbf{Traditional Methods}. (1) BPR\cite{BPR} is a classic matrix factorization method with a pair-wise ranking loss function. (2) ItemKNN\cite{itemknn} models item relations and use these relations to recommendations.
    \item \textbf{Cross-Domain Methods}. We select two classic CDR baselines: (1) NCF\cite{NCF} is a neural CF model that adopts multiple hidden layers to capture collaborative signals. We adapt it to the CDR problem by sharing user embeddings and setting domain-specific MLP networks for each domain. (2) CoNet\cite{hu2018conet} first models domain-specific interactions with MLP networks and then adopts cross-connection networks to transfer knowledge between domains.
    \item \textbf{Sequential Methods}. We include three classic deep sequential baselines based on network structures: RNN-based method GRU4Rec\cite{GRU4Rec}, Attention-based method SASRec\cite{SASRec}, and GNN-based method SR-GNN\cite{SRGNN}.
    \item \textbf{Cross-Domain Sequential Methods}. As the main competitors of the proposed method, we select CDSR baselines from two perspectives. \textit{Network-based}: (1) $\pi$-net\cite{pi-net} proposes a gating mechanism to transfer sequential patterns between domains. (2) MIFN\cite{MIFN} is a graph-based CDSR baseline, which utilizes knowledge graphs to guide the transfer process. \textit{SSL-based}: (1) C$^2$DSR\cite{c2dsr} proposes a contrastive infomax objective to capture and transfer user’s cross-domain sequential preferences. (2) Tri-CDR\cite{Tri-CDR} instead performs sequence-level contrastive learning to explore coarse-grained similarities and fine-grained distinctions among domain sequences.
\end{itemize}

\subsubsection{\textbf{Evaluation Protocols}}
We adopt the commonly used leave-one-out evaluation strategy to evaluate the top-K recommendation performance as in \cite{SASRec, bert4rec, s3-rec, c2dsr}. Specifically, we equally divide users' latest interaction sequences into validation and test sets, and all the remaining interactions are treated as the training set. As for evaluation metrics, we employ MRR\cite{MRR}, NDCG@10\cite{NDCG}, and HR@10. Besides, as suggested by Krichene and Rendle\cite{sampled_metrics}, we adopt the whole item set as the candidate item set during evaluation.

\subsubsection{\textbf{Implementation Details}}
We implement CA-CDSR based on PyTorch\cite{pytorch}. The maximum number of training epochs is set to 100. The training will be stopped when MRR on the validation data doesn't increase for 5 epochs. For CA-CDSR and all baselines, we adopt the following settings to perform a fair comparison: the training batch size $B$ is fixed to 256, Adam\cite{Adam} with a learning rate \{1e-4, 5e-4, 1e-3\} is used as the optimizer, the embedding size $d$ is set to 256 as previous works\cite{pi-net, c2dsr}, and the maximum sequence length $N$ is set to 30 for all datasets. Besides, the non-CDR methods are trained upon the mixed datasets.

Considering particular hyper-parameters of CA-CDSR, the number of GNN layers $L$ is fixed to 2, the strength of the two contrastive learning tasks $\lambda_1$ and $\lambda_2$ are searched from 0.1 to 0.9 with a step 0.1, the temperature $\tau$ of InfoNCE is fixed to 0.2, the magnitude of feature augmentation $\alpha$ is fixed to 0.1, the annealing step $N_{\text{anneal}}$ is searched among \{25, 50, 75, 100, 200\}.

\subsection{Overall Performance}
In this subsection, we compare the overall performance of all methods, which are presented in Table \ref{tab: overall performance}. Notably, the results of MIFN on “Entertain-Education” are missing, which is because we could not get the item knowledge graph. From the table, We can draw the following conclusions from the results: (1) Traditional methods yield inferior results compared to cross-domain methods, indicating the superiority of CDR methods in addressing the data sparsity issue. (2) CDSR methods, which consider both sequential characteristics and cross-domain modeling, achieve the best results across the three datasets. Additionally, methods based on self-supervised learning (SSL) demonstrate better performance, indicating the potential of contrastive learning in enhancing representation learning and knowledge transfer in CDSR. (3) Among the SSL-based CDSR methods, CA-CDSR consistently outperforms other approaches, particularly in the sparser datasets Food-Kitchen and Movie-Book. This can be attributed to two factors: Firstly, CA-CDSR adopts an SSL paradigm based on feature augmentation, which proves to be a more effective approach. Secondly, CA-CDSR incorporates representation alignment using carefully designed item representations, which guides the modeling of sequential user preferences and leads to improved performance compared to C$^2$DSR and Tri-CDR. (4) CA-CDSR exhibits superior performance on sparser datasets. We can observe more significant improvement in the two sparser datasets. This can be attributed to the fact that the limitations of previous self-supervised strategies have been mitigated by an ample supply of supervised signals present in the data.

\subsection{Ablation Studies}
\begin{table}[]
\caption{Abalation studies of CA-CDSR on MRR.}
\label{tab: ablation study}
\vspace{-0.2cm}
\resizebox{\linewidth}{!}{
\begin{tabular}{l|cc|cc|cc}
\hline
Datasets        & Food            & Kitchen         & Book            & Movie           & Entertain       & Education       \\ \hline
(A) CA-CDSR     & \textbf{0.0271} & \textbf{0.0105} & \textbf{0.0197} & \textbf{0.0104} & \textbf{0.4724} & \textbf{0.4714} \\ \hline
(B) -all        & 0.0239          & 0.0092          & 0.0163          & 0.0090          & 0.4566          & 0.4554          \\
(C) -generation & 0.0264          & 0.0102          & 0.0186          & 0.0099          & 0.4661          & 0.4664          \\
(D) -seq noise  & 0.0264          & 0.0103          & 0.0191          & 0.0101          & 0.4682          & 0.4695          \\
(E) -alignment  & 0.0259          & 0.0096          & 0.0182          & 0.0098          & 0.4625          & 0.4633          \\
(F) -ASF        & 0.0261          & 0.0098          & 0.0178          & 0.0098          & 0.4639          & 0.4621          \\
(G) -annealing  & 0.0266          & 0.0103          & 0.0191          & 0.0103          & 0.4699          & 0.4693          \\ \hline
\end{tabular}
}
\end{table}



In this subsection, we aim to validate the effectiveness of essential modules in CA-CDSR. The corresponding results are presented in Table \ref{tab: ablation study}, where several variations are examined: (B) removes both the representation generation and representation alignment by setting $\lambda_1$ and $\lambda_2$ to 0, (C) removes the representation generation by setting $\lambda_1$ to 0, (D) changes the sequence-aware noise distribution to the standard normal distribution $\mathcal{N}(\textbf{0}, \textbf{I})$, (E) removes the representation alignment by setting $\lambda_2$ to 0, (F) removes the ASF module, and (G) removes the annealing strategy by fixing $\eta$ to 0.5.


Comparing (A) to (B) shows significant performance degradation, highlighting the need to address partial item representation alignment in CDSR. Furthermore, comparing (A) to (C) and (E) reveals the contributions of representation modeling and alignment modules to overall performance. Comparing (A) to (D) shows performance drops due to inadequate feature augmentation based on random noise capturing essential sequential item correlations. Additionally, comparing (E) and (F) suggests that unfiltered full cross-domain alignment may even deteriorate performance. Comparing (A) to (F) demonstrates the significant performance enhancement from the ASF module, as it can filter out non-transferable knowledge during the alignment process. Lastly, comparing (A) to (G) shows slight improvement in recommendation performance with the annealing strategy for a more appropriate training process. Our empirical studies validate the effectiveness of each component in CA-CDSR.

\subsection{Further Experiments}

In this section, we aim to verify the generality of CA-CDSR, compare different augmentation strategies, and demonstrate the robustness of the hyper-parameters of CA-CDSR.

\subsubsection{\textbf{Generality of CA-CDSR}}\label{sec: versatility}

\begin{table}[]
\caption{Analysis of generality of CA-CDSR on MRR (\%).}
\label{tab: versatility study}

\resizebox{\linewidth}{!}{
\begin{tabular}{l|cc|cc|cc}
\hline
Datasets      & Food            & Kitchen         & Movie           & Book            & Entertain       & Education       \\ \hline
GRU4Rec       & 0.0143          & 0.0064          & 0.0088          & 0.0048          & 0.3252          & 0.3225          \\
GRU4Rec+      & \textbf{0.0194} & \textbf{0.0076} & \textbf{0.0115} & \textbf{0.007}  & \textbf{0.3564} & \textbf{0.3877} \\ \hline
SASRec        & 0.0176          & 0.0072          & 0.0091          & 0.0053          & 0.3523          & 0.3407          \\
SASRec+       & \textbf{0.0213} & \textbf{0.0076} & \textbf{0.0131} & \textbf{0.0081} & \textbf{0.3866} & \textbf{0.4091} \\ \hline
GNN-Attention & 0.0239          & 0.0092          & 0.0163          & 0.009           & 0.4566          & 0.4554          \\
CA-CDSR       & \textbf{0.0271} & \textbf{0.0105} & \textbf{0.0197} & \textbf{0.0104} & \textbf{0.4724} & \textbf{0.4714} \\ \hline
\end{tabular}
}
\end{table}


Since CA-CDSR is model-agnostic and can be easily combined with other sequential encoders, we have applied the proposed modules to some representative sequential encoders such as GRU4Rec \cite{GRU4Rec} and SASRec \cite{SASRec}. The corresponding results are summarized in Table \ref{tab: versatility study}, where '+' indicates the encoder enhanced by the proposed modules, and GNN-Attention refers to the graph-enhanced sequential encoder introduced in Sec. \ref{sec: sequential_modeling}. From the table, it can be observed that the performance of all encoders is greatly improved when combined with CA-CDSR, which further confirms the generality of CA-CDSR. Additionally, the results highlight the significance of item representation alignment in CDSR.

\subsubsection{\textbf{Comparison of different augmentation strategy}}

To verify the superiority of the proposed SAFA strategy, we compared it with other augmentation methods, including Edge Drop \cite{SGL} (graph structure augmentation), SimGCL \cite{SimGCL} (feature augmentation), and Tri-CDR \cite{Tri-CDR} (sequence-aware augmentation). The same sequential encoder is adopted for all methods. The results are presented in Table \ref{tab: augmentation comparison}, and we can draw these conclusions: (1) Comparing (A) with (B) and (C) with (D), the feature augmentation methods (B and D) outperform the non-sequential competitors (A and C), demonstrating the effectiveness of feature augmentation. (2) Comparing (A) with (C) and (B) with (D), the sequence-aware augmentation strategies (C and D) outperform the non-sequential competitors (A and B). This indicates that sequence-aware strategies can capture essential sequential information, leading to improved performance.

\begin{table}[]
\caption{Results of augmentation strategies on MRR (\%).}
\label{tab: augmentation comparison}

\renewcommand\arraystretch{0.88}
\begin{tabular}{l|cc|cc}
\hline
Datasets         & Food            & Kitchen         & Movie           & Book            \\ \hline
Sequence Encoder & 0.0239          & 0.0092          & 0.0163          & 0.0090          \\ \hline
(A) Edge Drop    & 0.0232          & 0.0086          & 0.0168          & 0.0085          \\
(B) SimGCL       & 0.0244          & 0.0094          & 0.0168          & 0.009           \\ \hline
(C) Tri-CDR      & 0.0254          & 0.0098          & 0.0179          & 0.0098          \\
(D) CA-CDSR      & \textbf{0.0271} & \textbf{0.0105} & \textbf{0.0197} & \textbf{0.0104} \\ \hline
\end{tabular}
\end{table}

\subsubsection{\textbf{Hyper-Parameter Sensitivity}}

In this section, we investigate the sensitivity of hyper-parameters in CA-CDSR, specifically the strength of representation modeling ($\lambda_1$) and representation alignment ($\lambda_2$) in Eq. \ref{eq: total loss}. Since the trend is similar for the three datasets, we only report the results for the Food-Kitchen dataset.

\begin{figure}[t!]
    \centering
    \begin{subfigure}[t]{0.22\textwidth}
           \centering
           \includegraphics[width=\textwidth]{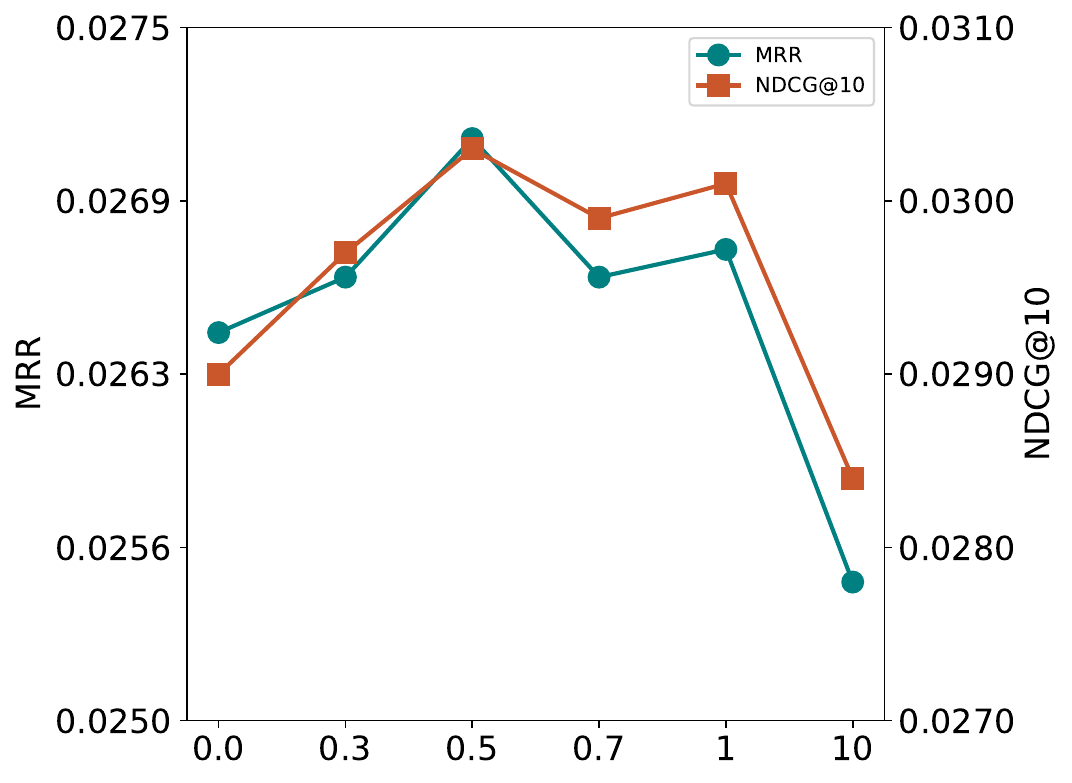}
            \caption{Food}
            \label{fig: lambda1_food}
    \end{subfigure}
    \begin{subfigure}[t]{0.22\textwidth}
            \centering
            \includegraphics[width=\textwidth]{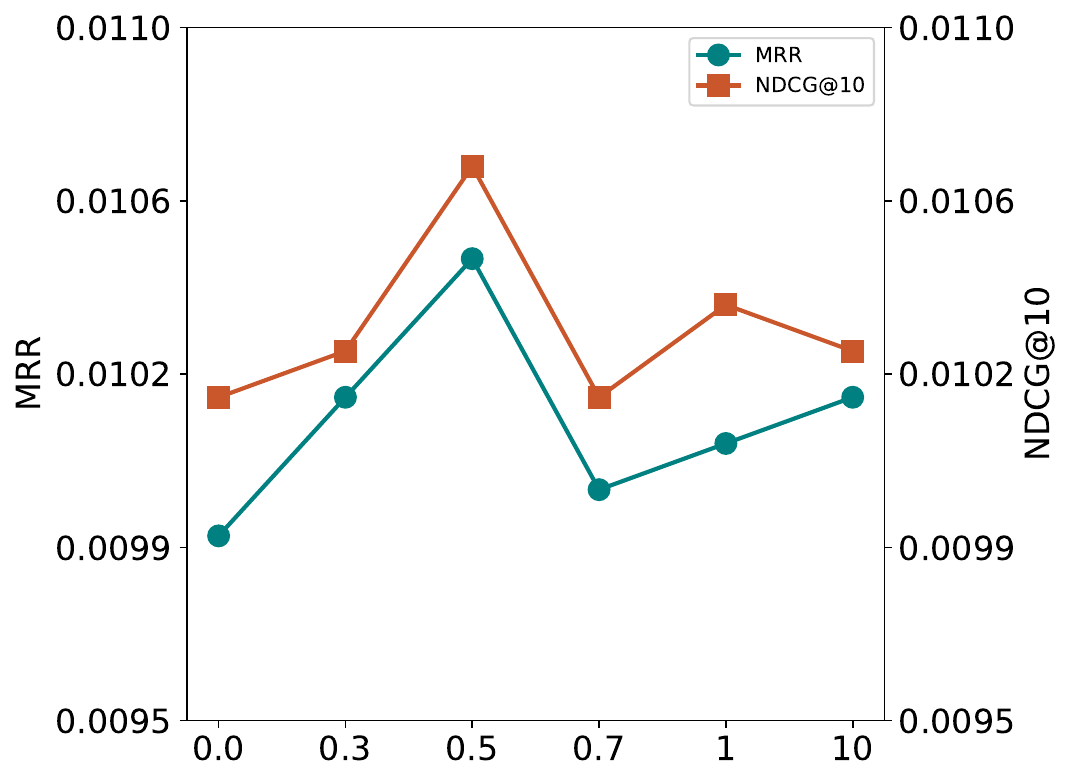}
            \caption{Kitchen}
            \label{fig: lambda1_kitchen}
    \end{subfigure}

    \caption{Recommendation performance (\%) w.r.t $\lambda_1$.}
    \label{fig: lambda1 analysis}    

\end{figure}

We set $\lambda_1$ in the range of $[0, 0.3, 0.5, 0.7, 1, 10]$ and the results are depicted in Fig. \ref{fig: lambda1 analysis}. In the figure, the performance exhibits an increasing trend as $\lambda_1$ increases, as modeling collaborative and sequential item correlations can facilitate the representation alignment process. However, the performance decreases when $\lambda_1$ exceeds a threshold, as too large $\lambda_1$ will instead suppress the representation alignment. 

\begin{figure}[t!]
    \centering
    \begin{subfigure}[t]{0.22\textwidth}
           \centering
           \includegraphics[width=\textwidth]{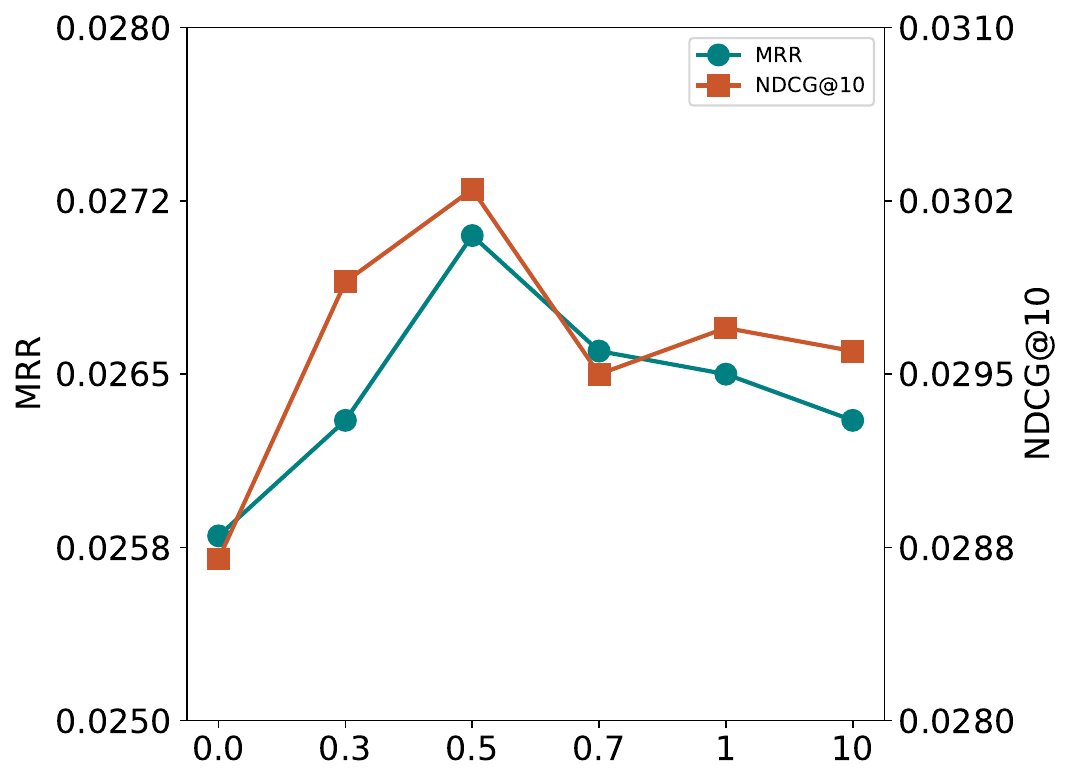}
            \caption{Food}
            \label{fig: lambda2_food}
    \end{subfigure}
    \begin{subfigure}[t]{0.22\textwidth}
            \centering
            \includegraphics[width=\textwidth]{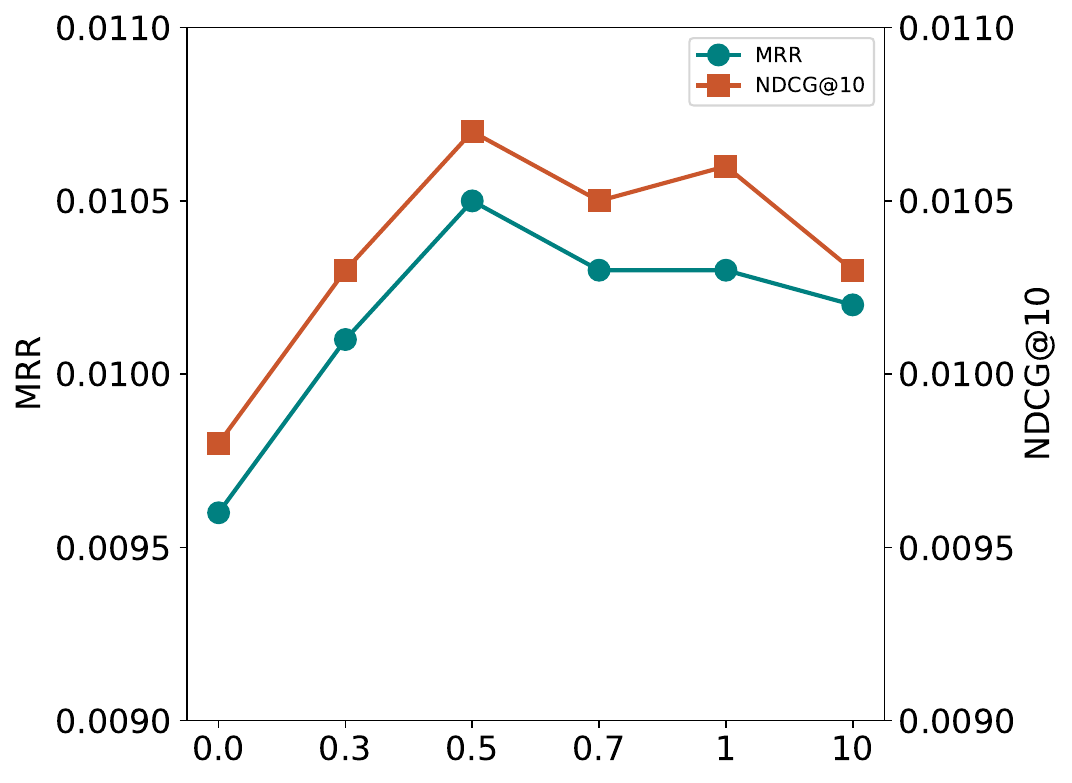}
            \caption{Kitchen}
            \label{fig: lambda2_kitchen}
    \end{subfigure}
    \vspace{-0.3cm}

    \caption{Recommendation performance (\%) w.r.t $\lambda_2$.}
    \label{fig: lambda2 analysis}
    \vspace{-0.3cm}
\end{figure}


We set $\lambda_2$ in the range of $[0, 0.3, 0.5, 0.7, 1, 10]$, and the results are illustrated in Fig. \ref{fig: lambda2 analysis}. The performance initially increases and then decreases. This observation suggests that the domain gap should be narrowed to an optimal extent. Insufficient representation alignment may not fully exploit the cross-domain knowledge, while excessive representation alignment can impair the inherent gap between domains, resulting in the negative transfer problem.

\section{CONCLUSIONS}
In this paper, we proposed a model-agnostic framework named CA-CDSR, which studies cross-domain sequential recommendation from a partial item representation alignment perspective. Particularly, we proposed a sequence-aware feature augmentation strategy and an adaptive spectrum filter, which respectively achieved sequence-aware generation and adaptive partial alignment for item representations. Extensive experiments demonstrated that CA-CDSR not only outperformed other baselines with significant margins but also exhibited promising potential in aligning items in the representation space. In the future, we will explore the partial alignment problem on datasets with accessible item properties, which can provide explicit guidance for the alignment process.

\begin{acks}

\end{acks}

\bibliographystyle{ACM-Reference-Format}
\bibliography{content/references}


\end{document}